\documentclass{article}
\usepackage{graphicx}
\usepackage{amsmath}

\input{tcilatex}

\begin{document}

\title{\textbf{Fundamental Commutators in a}\\
\textbf{Gravitational Field }}
\author{W. Chagas-Filho \\
Departamento de Fisica, Universidade Federal de Sergipe\\
SE, Brazil}
\maketitle

\begin{abstract}
We show how an induced invariance of the massless particle action can be
used to construct an extension of the Heisenberg canonical commutation
relations in a non-commutative space-time.
\end{abstract}

\section{Introduction}

\noindent The issue of non-commutativity of the space-time coordinates has
become a central one in all discussions about the description of physics at
the Planck scale. The large amounts of momentum and energy necessary to
probe the physics at this length scale must necessarily modify the
space-time geometry, and this modification affects the outcome of quantum
measurement processes, introducing additional new uncertainties among the
canonical variables. In particular, position measurements are expected to
non commute. A rigorous way to find and incorporate these additional new
uncertainties in a definite physical theory is presently unknown. The issue
of non-commutative space-time coordinates also makes its appearance in
discussions concerning formulations of quantum gravity. Quantum gravity has
an uncertainty principle [1] which prevents one from measuring positions to
better accuracies than the Planck length. This effect would then be modeled
by a non-vanishing commutation relation between the space-time coordinates.
In these two related [2] situations space-time non-commutativity introduces
non-locality, and it is far from clear that Poincar\'{e} invariance can
survive in a non-local, non-commutative, space-time geometry.

The point of view we explore in this work is that non-locality brings with
it deep conceptual issues which have not yet been well understood. It is
therefore useful to try to understand these issues in the simplest examples
first, before proceeding to a more realistic theory of quantum gravity. The
simplest example of a generally covariant theory is relativistic particle
theory. The relativistic particle action is therefore the simplest
theoretical laboratory where to search for the dynamical origin of a
non-commutative space-time geometry, which is what ultimately causes the
non-locality.

The ADM construction [3] of general relativity made it clear that the mass
of a particle comes from its interactions with fields other than
gravitational. A massless particle couples only to the gravitational field.
Because of the existence of the holographic principle, massless particles
also play important roles in the description of the low and high energy
physics of M-theory [4]. By considering the dynamics of massless particles,
we can access at least some part of the full physics of M-theory. For these
reasons, in this work we concentrate only in the massless sector of
relativistic particle theory. Our contribution to the literature is to
present a theoretical path to new canonical commutation relations which
incorporate uncertainties arising from gravitational effects. We find here
that we can use the special-relativistic orthogonality condition between the
velocity and the acceleration to induce the appearance of a new type of
local scale invariance in the massless particle action functional. In the
transition to the Hamiltonian formalism this new scale invariance causes the
appearance of new canonical commutator relations between the transformed
canonical variables. These new commutators obey all Jacobi identities among
the canonical variables and preserve the structure of the Poincar\'{e}
algebra. The results contained in this work are therefore important for the
following two reasons. First, they make it evident that the Heisenberg
canonical commutator relations are not the adequate canonical commutator
relations to be used in quantum mechanical calculations when strong
gravitational fields are present. Second, they make it evident that the
adequate canonical commutator relations preserve the structure of the
Poincar\'{e} space-time algebra. These observations can perhaps, in the
future, solve the problem of the apparent incompatibility between general
relativity and quantum mechanics.

\section{Relativistic Particles}

A relativistic particle describes is space-time a one-parameter trajectory $%
x^{\mu }(\tau )$. A possible form of the action is the one proportional to
the arc length traveled by the particle and given by 
\begin{equation}
S=-m\int ds=-m\int d\tau \sqrt{-\dot{x}^{2}}  \tag{2.1}
\end{equation}
In this work we choose $\tau $ to be the particle's proper time, $m$ is the
particle's mass and $ds^{2}=-\delta _{\mu \nu }dx^{\mu }dx^{\nu }$. We work
in an Euclidean space-time and so the index $\mu $ takes the values 1,2,3,4,
with $x^{4}=ict$. A dot denotes derivatives with respect to $\tau $ and we
use units in which $\hbar =c=1$.

Action (2.1) is invariant under the Poincar\'{e} transformation 
\begin{equation}
\delta x^{\mu }=a^{\mu }+\omega _{\nu }^{\mu }x^{\nu }  \tag{2.2}
\end{equation}
where $a^{\mu }$ is a constant vector and $\omega _{\mu \nu }=-\omega _{\nu
\mu }$. As a consequence of the invariance of action (2.1) under
transformation (2.2), the following vector field can be defined in
space-time 
\begin{equation}
V=ia^{\mu }p_{\mu }-\frac{i}{2}\omega ^{\mu \nu }M_{\mu \nu }  \tag{2.3}
\end{equation}
where $p_{\mu }=-i\partial _{\mu }$ and $M_{\mu \nu }=x_{\mu }p_{\nu
}-x_{\nu }p_{\mu }$. Introducing the fundamental Heisenberg commutators 
\begin{equation}
\lbrack p_{\mu },p_{\nu }]=0  \tag{2.4a}
\end{equation}
\begin{equation}
\lbrack x_{\mu },p_{\nu }]=i\delta _{\mu \nu }  \tag{2.4b}
\end{equation}
\begin{equation}
\lbrack x_{\mu },x_{\nu }]=0  \tag{2.4c}
\end{equation}
we find that the generators of the vector field $V$ obey the algebra 
\begin{equation}
\lbrack p_{\mu },p_{\nu }]=0  \tag{2.5a}
\end{equation}
\begin{equation}
\lbrack p_{\mu },M_{\nu \lambda }]=i\delta _{\mu \nu }p_{\lambda }-i\delta
_{\mu \lambda }p_{\nu }  \tag{2.5b}
\end{equation}
\begin{equation}
\lbrack M_{\mu \nu },M_{\rho \lambda }]=i\delta _{\nu \lambda }M_{\mu \rho
}+i\delta _{\mu \rho }M_{\nu \lambda }-i\delta _{\nu \rho }M_{\mu \lambda
}-i\delta _{\mu \lambda }M_{\nu \rho }  \tag{2.5c}
\end{equation}
This is the Poincar\'{e} space-time algebra. Action (2.1) is also invariant
under the reparametrization 
\begin{equation}
\tau \rightarrow \tau ^{\prime }=f(\tau )  \tag{2.6}
\end{equation}
where $f$ is an arbitrary continuous function of $\tau .$ As a consequence
of its invariance under transformation (2.6), the particle action (2.1)
defines the simplest possible generally covariant physical system.

Action (2.1) is obviously inadequate to study the massless limit of particle
theory and so we must find an alternative action. Such an action can be
easily computed by treating the relativistic particle as a constrained
system. In the transition to the Hamiltonian formalism action (2.1) gives
the canonical momentum 
\begin{equation}
p_{\mu }=\frac{m}{\sqrt{-\dot{x}^{2}}}\dot{x}_{\mu }  \tag{2.7}
\end{equation}
and this momentum gives rise to the primary constraint 
\begin{equation}
\phi =\frac{1}{2}(p^{2}+m^{2})=0  \tag{2.8}
\end{equation}
In this work we follow Dirac's [5] convention that a constraint is set equal
to zero only after all calculations have been performed. The canonical
Hamiltonian corresponding to action (2.1), $H=p.\dot{x}-L$, identically
vanishes. Dirac's Hamiltonian for the relativistic particle is then 
\begin{equation}
H_{D}=H+\lambda \phi =\frac{1}{2}\lambda (p^{2}+m^{2})  \tag{2.9}
\end{equation}
where $\lambda (\tau )$ is a Lagrange multiplier, to be interpreted as an
independent variable. The Lagrangian that corresponds to (2.9) is 
\begin{equation}
L=p.\dot{x}-\frac{1}{2}\lambda (p^{2}+m^{2})  \tag{2.10}
\end{equation}
Solving the equation of motion for $\ p_{\mu }$ that \ follows from (2.10)
and inserting the result back in it, we obtain the particle action 
\begin{equation}
S=\int d\tau (\frac{1}{2}\lambda ^{-1}\dot{x}^{2}-\frac{1}{2}\lambda m^{2}) 
\tag{2.11}
\end{equation}
In action (2.11), $\lambda (\tau )$ can be associated [6] to a ``world-line
metric'' $\gamma _{\tau \tau }$ , $\lambda (\tau )=[-\gamma _{\tau \tau
}(\tau )]^{\frac{1}{2}}$ , such that $ds^{2}=\gamma _{\tau \tau }d\tau d\tau 
$. In (2.11), $\lambda (\tau )$ is an ``einbein'' field. In more dimensions,
the ``vielbein'' $e_{\mu }^{a}$ is an alternative description of the metric
tensor [4]. In this context, the particle mass $m$ plays the role of a
(0+1)-dimensional ``cosmological constant''. Action (2.11) is classically
equivalent to action (2.1). This can be checked in the following way. If we
solve the classical equation of motion for $\lambda (\tau )$ that follows
from (2.11) we get the result $\lambda =\pm (\sqrt{-\dot{x}^{2}}/m)$.
Inserting the solution with the positive sign in (2.11), it becomes
identical to (2.1).\ The great advantage of action (2.11) is that it has a
smooth transition to the $m=0$ limit.

The general covariance of action (2.11) manifests itself through invariance
under the transformation 
\begin{equation}
\delta x^{\mu }=\epsilon \dot{x}^{\mu }  \tag{2.12a}
\end{equation}
\begin{equation}
\delta \lambda =\frac{d}{d\tau }(\epsilon \lambda )  \tag{2.12b}
\end{equation}
where $\epsilon (\tau )$ is an arbitrary infinitesimal parameter. Varying $%
x^{\mu }$ in (2.11) we obtain the classical equation for free motion 
\begin{equation}
\frac{d}{d\tau }(\frac{\dot{x}_{\mu }}{\lambda })=\frac{dp_{\mu }}{d\tau }=0
\tag{2.13}
\end{equation}

Now we make a transition to the massless limit. This limit is described by
the action 
\begin{equation}
S=\frac{1}{2}\int d\tau \lambda ^{-1}\dot{x}^{2}  \tag{2.14}
\end{equation}
Action (2.14) is invariant under the Poincar\'{e} transformation (2.2) with $%
\delta \lambda =0$, and under the infinitesimal reparametrization (2.12).
The classical equation of motion for $x^{\mu }$ that follows from action
(2.14) is identical to (2.13). The equation of motion for $\lambda $ gives
the condition $\dot{x}^{2}=0$, which tells us that a massless particle moves
at the speed of light. As a consequence of this, it becomes impossible to
solve for $\lambda (\tau )$ from its equation of motion. In the massless
theory the value of $\lambda (\tau )$ is completely arbitrary.

In the transition to the Hamiltonian formalism action (2.14) gives the
canonical momenta 
\begin{equation}
p_{\lambda }=0  \tag{2.15}
\end{equation}
\begin{equation}
p_{\mu }=\frac{\dot{x}_{\mu }}{\lambda }  \tag{2.16}
\end{equation}
and the canonical Hamiltonian 
\begin{equation}
H=\frac{1}{2}\lambda p^{2}  \tag{2.17}
\end{equation}
Equation (2.15) is a primary constraint. Introducing the Lagrange multiplier 
$\xi (\tau )$ for this constraint we can write the Dirac Hamiltonian 
\begin{equation}
H_{D}=\frac{1}{2}\lambda p^{2}+\xi p_{\lambda }  \tag{2.18}
\end{equation}
Requiring the dynamical stability of constraint (2.15), $\dot{p}_{\lambda
}=\{p_{\lambda },H_{D}\}=0$, we obtain the secondary constraint 
\begin{equation}
\phi =\frac{1}{2}p^{2}=0  \tag{2.19}
\end{equation}

Now we may use the fact that we are dealing with a special-relativistic
system. Special relativity has the dynamical feature [7] that the
relativistic velocity is always orthogonal to the relativistic acceleration, 
$\dot{x}.\ddot{x}=0$. We can use this orthogonality to induce the invariance
of the massless particle action (2.14) under the transformation 
\begin{equation}
x^{\mu }\rightarrow \tilde{x}^{\mu }=\exp \{\beta (\dot{x}^{2})\}x^{\mu } 
\tag{2.20a}
\end{equation}
\begin{equation}
\lambda \rightarrow \exp \{2\beta (\dot{x}^{2})\}\lambda  \tag{2.20b}
\end{equation}
where $\beta $ is an arbitrary function of $\dot{x}^{2}$. Transformation
(2.20) is a new type of local scale invariance of the action for a massless
particle. Action (2.14) is invariant under (2.20) because the orthogonality
condition forces $\dot{x}^{\mu }$ to transform as $\exp \{\beta (\dot{x}%
^{2})\}\dot{x}^{\mu }$ when $x^{\mu }$ transforms as in (2.20a). We
emphasize that although the orthogonality condition must be used to get the
invariance of action (2.14) under transformation (2.20), this condition is
not an external ingredient in the theory. In fact, the orthogonality between
the relativistic velocity and acceleration is an unavoidable condition here.
It is an imposition of special relativity.

Consider now the commutator structure that transformation (2.20a) induces on
the canonical operators. From the definition (2.16) of the canonical
momentum we find that the momenta must transform as 
\begin{equation}
p_{\mu }\rightarrow \tilde{p}_{\mu }=\exp \{-\beta (\dot{x}^{2})\}p_{\mu } 
\tag{2.21}
\end{equation}
when $x^{\mu }$ transforms as in (2.20a). It can be verified that
transformations (2.20) and (2.21) together leave invariant the $m=0$ limit
of the first order Lagrangian (2.10), and the massless particle Hamiltonian
(2.17). These observations confirm that transformation (2.20) is a true
invariance of the massless particle action (2.14).

Taking $\beta (\dot{x}^{2})=\beta (\lambda ^{2}p^{2})$ in transformations
(2.20a) and (2.21), and retaining only the linear terms in $\beta $ in the
exponentials, we find that the new transformed canonical variables $(\tilde{x%
}_{\mu },\tilde{p}_{\mu })$ obey the commutators 
\begin{equation}
\lbrack \tilde{p}_{\mu },\tilde{p}_{\nu }]=0  \tag{2.22a}
\end{equation}
\begin{equation}
\lbrack \tilde{x}_{\mu },\tilde{p}_{\nu }]=(1+\beta )\{i\delta _{\mu \nu
}(1-\beta )-[x_{\mu },\beta ]p_{\nu }\}  \tag{2.22b}
\end{equation}
\begin{equation}
\lbrack \tilde{x}_{\mu },\tilde{x}_{\nu }]=(1+\beta )\{x_{\mu }[\beta
,x_{\nu }]-x_{\nu }[\beta ,x_{\mu }]\}  \tag{2.22c}
\end{equation}
written in terms of the old canonical variables. These commutators obey the
non trivial Jacobi identities $(\tilde{x}_{\mu },\tilde{x}_{\nu },\tilde{x}%
_{\lambda })=0$ \ and \ $(\tilde{x}_{\mu },\tilde{x}_{\nu },\tilde{p}%
_{\lambda })=0$. They also reduce to the usual Heisenberg commutators when
constraint (2.19) is imposed. We see from commutator (2.22c) that
transformation (2.20a) induces a transition to a non-commutative space-time
geometry.

The non-commutative space-time geometry is completely determined by the
choice $\beta (\dot{x}^{2})=\beta (\lambda ^{2}p^{2})=\lambda ^{2}p^{2}$.
This is because commutators satisfy the property $[A^{n},B]=nA^{n-1}[A,B]$.
If we consider, for instance, $\beta =(\lambda ^{2}p^{2})^{2}$ and compute $%
[\beta ,x^{\mu }]$ we will find $[\beta ,x^{\mu }]=2\lambda
^{2}p^{2}[\lambda ^{2}p^{2},x^{\mu }]$, and this vanishes when constraint
(2.19) is imposed. Similarly, all higher order terms will vanish when (2.19)
is imposed, and the space-time geometry is completely determined by the case 
$\beta =\lambda ^{2}p^{2}$. Computing the commutators (2.22b) and (2.22c)
for this form of $\beta ,$ \textbf{and finally imposing constraint }(2.19),
we arrive at the canonical commutation relations 
\begin{equation}
\lbrack \tilde{p}_{\mu },\tilde{p}_{\nu }]=0  \tag{2.23a}
\end{equation}
\begin{equation}
\lbrack \tilde{x}_{\mu },\tilde{p}_{\nu }]=i\delta _{\mu \nu }-i\lambda
^{2}p_{\mu }p_{\nu }  \tag{2.23b}
\end{equation}
\begin{equation}
\lbrack \tilde{x}_{\mu },\tilde{x}_{\nu }]=-2i\lambda ^{2}(x_{\mu }p_{\nu
}-x_{\nu }p_{\mu })  \tag{2.23c}
\end{equation}
while the transformation equations (2.20a) and (2.21) become 
\begin{equation}
x^{\mu }\rightarrow \tilde{x}^{\mu }=x^{\mu }  \tag{2.24a}
\end{equation}
\begin{equation}
p_{\mu }\rightarrow \tilde{p}_{\mu }=p_{\mu }  \tag{2.24b}
\end{equation}
We can then write down the canonical commutators 
\begin{equation}
\lbrack p_{\mu },p_{\nu }]=0  \tag{2.25a}
\end{equation}
\begin{equation}
\lbrack x_{\mu },p_{\nu }]=i\delta _{\mu \nu }-i\lambda ^{2}p_{\mu }p_{\nu }
\tag{2.25b}
\end{equation}
\begin{equation}
\lbrack x_{\mu },x_{\nu }]=-2i\lambda ^{2}(x_{\mu }p_{\nu }-x_{\nu }p_{\mu })
\tag{2.25c}
\end{equation}
Notice that, after imposing constraint (2.19), we can not return to a
commutative space-time by performing the inverse transformation because the
transformations (2.20a) and (2.21) become the identity transformation, as
given by (2.24). Now the only way to return to a commutative space-time is
with the vanishing of the (0+1)-dimensional gravitational field described by 
$\lambda (\tau )$. This makes it evident that the Heisenberg canonical
commutation relations (2.4) are adequate only in the presence of vanishing,
or at best very weak, gravitational fields.

Commutators (2.25) obey all Jacobi identities among the canonical variables.
It can be also verified that if we compute the algebra of the generators of
the vector field (2.3) using the commutators (2.25), instead of the
Heisenberg commutators (2.4), the same expressions (2.5) are reproduced.
Commutators (2.25) preserve the structure of the Poincar\'{e} space-time
algebra. Since in the massless theory the value of $\lambda (\tau )$ is
arbitrary, as we saw above, we can use the reparametrization invariance
(2.12) to choose a gauge in which $\lambda =1$. We then end up with the
convenient commutators 
\begin{equation}
\lbrack p_{\mu },p_{\nu }]=0  \tag{2.26a}
\end{equation}
\begin{equation}
\lbrack x_{\mu },p_{\nu }]=i\delta _{\mu \nu }-ip_{\mu }p_{\nu }  \tag{2.26b}
\end{equation}
\begin{equation}
\lbrack x_{\mu },x_{\nu }]=-2i(x_{\mu }p_{\nu }-x_{\nu }p_{\mu }) 
\tag{2.26c}
\end{equation}
Commutators (2.26) form a consistent set of quantum commutators on which an
entirely new formulation of quantum mechanics in a gravitational field can
be based. Perhaps this new formulation can have the power to bring the
apparent incompatibility between general relativity and quantum mechanics to
an end.

\section{Conclusion}

In this work we deduced a new set of canonical commutation relations for the
massless relativistic particle. This new set generalizes the usual
Heisenberg canonical commutators to the case when a gravitational field is
present. The origin of these new commutators is a new type of local scale
invariance of the massless particle action. This new invariance manifests
itself when the special-relativistic orthogonality condition between the
velocity and the acceleration is imposed. The new commutators obey all
Jacobi identities among the canonical variables and, although they bring in
a certain amount of non-locality, this non-locality does not conflict with
the Poincar\'{e} invariance of the theory because the new commutators leave
the Poincar\'{e} space-time algebra invariant. The new commutators therefore
form a consistent set of quantum commutators which could solve the apparent
incompatibility between general relativity and quantum mechanics.


\begin{thebibliography}{9}
\bibitem{1}  \noindent B. DeWitt, \textsl{Gravitation}, edited by L. Witten,
1962

\bibitem{2}  D. V. Ahluwalia, \textsl{Quantum measurement, gravitation, and
locality}, Phys. Lett. B339 (1994) 301 , (gr-qc/9308007)

\bibitem{3}  R. Arnowitt, S. Deser and C. W. Misner, Phys. Rev. 117 (1960)
1595; ibid \textsl{The dynamics of general relativity, }(gr-qc/0405109)

\bibitem{4}  L. Susskind, \textsl{The world as a hologram,} J. Math. Phys.
36 (1995) 6377 (hep-th/9409089); T. Banks, W. Fischler, S. H. Shenker and L.
Susskind, \textsl{M theory as a matrix model: a conjecture}; Phys. Rev. D55
(1997) 5112 (hep-th/9610043); W. Taylor, \textsl{M(atrix) theory: matrix
quantum mechanics as a fundamental theory}, Rev. Mod. Phys. 73 (2001) 419
(hep-th/0101126)

\bibitem{5}  P. A. M. Dirac, \textsl{Lectures on Quantum Mechanics, }Yeshiva
University,1964

\bibitem{6}  C. V. Johnson, \textsl{D-branes}, Cambridge Monographs on
Mathematical Physics, 2003

\bibitem{7}  L. Landau and L. Lifchitz, \textsl{\ Theorie du Champ, }%
Editions MIR, Moscow, 1966
\end{thebibliography}
\end{document}